\documentclass[twocolumn,showpacs,preprintnumbers,amsmath,amssymb]{revtex4}
\usepackage{graphicx}% Include figure files
\usepackage{dcolumn}% Align table columns on decimal point
\usepackage{bm}% bold math

\begin{document}

\preprint{APS/123-QED}

\title{Magneto-optical trap for metastable helium at 389 nm}
\author{J.C.J. Koelemeij}
\email{koel@nat.vu.nl}
\author{R.J.W. Stas}
\author{W. Hogervorst}
\author{W. Vassen}
\affiliation{Laser Centre Vrije Universiteit, De Boelelaan 1081, 1081 HV Amsterdam, The Netherlands}

\date{\today}% It is always \today, today,

\begin{abstract}
We have constructed a magneto-optical trap (MOT) for metastable
triplet helium atoms utilizing the $2 \text{ }^3S_1 \rightarrow 3
\text{ }^3P_2$ line at 389 nm as the trapping and cooling
transition. The far-red-detuned MOT (detuning $\Delta = - 41\text{
MHz}$) typically contains few times $10^7$ atoms at a relatively 
high ($\sim 10^9$ cm$^{-3}$) density,
which is a consequence of the large momentum transfer per photon at
389 nm and a small two-body loss rate coefficient ($2\times 10^{-10}
\text{ cm}^3/\text{s} < \beta < 1.0\times 10^{-9}\text{
cm}^3/\text{s}$). The two-body loss rate is more than five
times smaller than in a MOT on the commonly used $2 \text{ }^3S_1
\rightarrow 2 \text{ }^3P_2$ line at 1083 nm. Furthermore, we
measure a temperature of 0.46(1) mK, a factor 2.5 lower as compared
to the 1083 nm case. Decreasing the detuning to $\Delta=-9$ MHz
results in a cloud temperature as low as 0.25(1) mK, at small number
of trapped atoms. The 389 nm MOT exhibits small losses
due to two-photon ionization, which have been investigated as well.

\end{abstract}

\maketitle

\section{\label{introduction}Introduction}
A magneto-optical trap (MOT) is a standard tool in the production of
cold atomic gases, allowing investigation of cold-collision
phenomena~\cite{Weiner} as well as the realization of Bose-Einstein
condensation (BEC) in alkali species~\cite{Anderson} and, more
recently, in metastable triplet helium
(He*)~\cite{OrsayBEC,ENSBEC}. He* has a high (19.8 eV) internal
energy, which allows for real-time diagnostics and increased
sensitivity in BEC probing. Unfortunately, the high internal energy
also introduces strong Penning ionization losses in magneto-optically
trapped atomic clouds, which imposes limits on the maximum
achievable density. The two-body loss rate coefficient related to
this process is about $5\times 10^{-9}\text{ cm}^3/\text{s}$ for a MOT
on the $2 \text{ }^3S_1 \rightarrow 2 \text{ }^3P_2$ transition at
1083 nm~\cite{Tol}, which is about two orders of magnitude larger than
the loss rate coefficient in a standard alkali MOT. In BEC
experiments, the MOT is used as a bright source of cold atoms to load
a magnetic trap with large numbers of atoms. Moreover, as a starting
point for evaporative cooling a dense magnetostatically trapped cloud
is desired. So ideally, the magneto-optically trapped cloud must
provide this high density. In the present work, we explore
the feasibility and possible advantages of a MOT using the $2\text{
}^3S_1 \rightarrow 3\text{ }^3P_2$ transition at 389 nm for metastable
helium, in comparison with the conventional $2\text{ }^3S_1
\rightarrow 2\text{ }^3P_2$ (1083 nm) magneto-optical trap.

\begin{figure}
\includegraphics[scale=1]{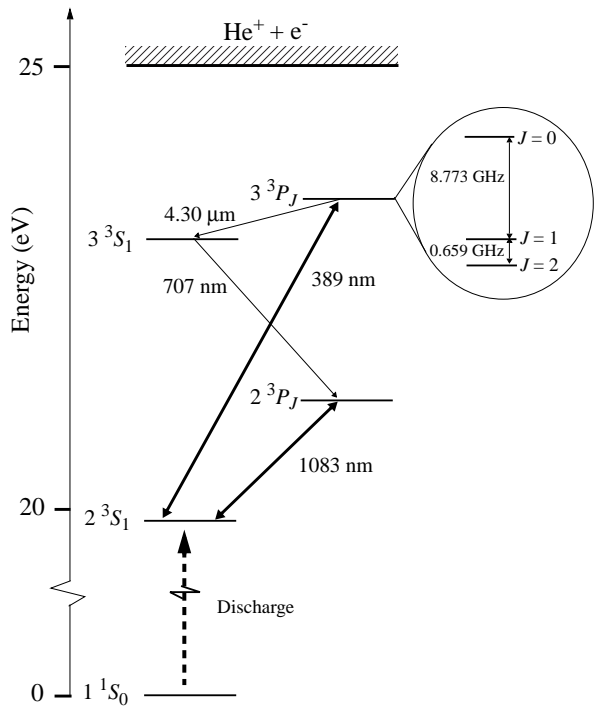}
\caption{\label{levelscheme}Helium level scheme. The long-lived 
2 $^3S_1$ metastable state is populated in a DC discharge. The
$2\text{ }^3S_1
\rightarrow 3\text{ }^3P_2$ (389 nm) and $2\text{ }^3S_1
\rightarrow 2\text{ }^3P_2$ (1083 nm) laser cooling transitions are indicated
with bold arrows.}
\end{figure}

Although the 389 nm transition was used recently in laser cooling
experiments~\cite{Schumann}, it has not found wide application yet.
This mainly relates to the fact that $10 \%$ of the $3\text{ }^3P_2$
population decays via the $3\text{ }^3S_1$ state
(Fig.~\ref{levelscheme}), making a closed laser cooling transition
between magnetic substates impossible. In addition, the shorter 389 nm
wavelength leads, in combination with a linewidth $\Gamma / 2\pi = 1.5
\text{ MHz}$~\cite{Drake}, to a relatively high saturation intensity
$I_0=3.31 \text{ mW/cm}^2$ (circular polarization in an optically
pumped environment~\cite{MetcalfStraten}). In comparison, the $2\text{
}^3S_1\rightarrow 2\text{ }^3P_2$ transition at 1083 nm has almost the
same linewidth but a saturation intensity of only $0.17\text{ mW/cm}^2$. To
maximize the number of trapped atoms, dedicated metastable helium
magneto-optical traps are operated at large detuning and
intensity~\cite{Tol,OrsayMOT,ENSpobec}.
This implies the need for a high-power laser
setup. Nevertheless, the concept of a 389 nm magneto-optical trap is
appealing. An attractive feature of the 389 nm transition is the
momentum transfer per photon, which is 2.8 times larger than for 1083
nm photons. Since both transitions have nearly equal linewidths, the
spontaneous cooling force increases proportional to the photon
momentum. This opens the possibility to compress the cloud
substantially in comparison to a 1083 nm MOT at the same detuning and
power. Unfortunately, compression may lead to increased losses
predominantly due to light-assisted two-body collisions. The two-body loss
rate coefficient for the 389 nm situation is, however, unknown. In
case of a relatively low rate coefficient, the cloud may be compressed
without loss of too many metastables. Furthermore, it should be noted
that the 1083 nm and 389 nm transitions are electronically alike,
which greatly facilitates the comparison between the two MOT
types. Finally, the 389 nm MOT differs from the 1083 nm MOT in yet
another respect: two 389 nm photons contain sufficient energy to
ionize an atom in the $2 \text{ } ^3S_1$ state. This may introduce
observable additional losses.

In this article, we report on the study of a prototype 389 nm MOT.
In Sec.~\ref{MOTloading} we present some typical aspects of the 389 nm MOT from
general laser cooling theory. Next, we outline our experimental setup in 
Sec.~\ref{experimental}.
Results are given in Sec.~\ref{results}. Conclusive remarks and an outlook are
presented in Sec.~\ref{conclusion}.
\section{\label{MOTloading}Qualitative description of the 389 nm MOT}
\subsection{\label{springdamping}Spring constant and damping coefficient}
The large Doppler cooling force modifies the equilibrium conditions in a 389
nm MOT with respect to the 1083 nm situation. This follows from regarding 
the motion of an atom, trapped in a one-dimensional MOT, as an overdamped
harmonic oscillation~\cite{Lett,Sengstock}. The oscillation frequency
$\omega_{osc}$ and damping coefficient $\epsilon_d$ are, for small
velocities and small deviations from trap center, given by
\begin{eqnarray}
\omega_{osc}^2&=&4 \hbar k \frac{4\delta S \zeta}{m(1+2S+4
\delta^2)^2}\label{omega},
\\
\epsilon_d&=&4 \hbar k^2 \frac{4\delta S}{m(1+2S+4 \delta^2)^2}\label{epsilon},
\end{eqnarray}
with $k$ the wavenumber of the MOT laser light, $m$ the atomic mass,
$\delta=2\pi \Delta/\Gamma$ with $\Delta$ the laser detuning from
resonance in MHz, $S=I/I_0$ the saturation parameter, with $I$ the
intensity per MOT beam, and $\zeta$ representing the spatial derivative of
the location-dependent Zeeman detuning. The harmonic oscillator
frequency is related to the spring constant $\kappa$ via
$\omega_{osc}^2=\kappa/m$. Two general differences between the 389 nm
and 1083 nm MOT follow immediately from Eq.~(\ref{omega}) and
Eq.~(\ref{epsilon}). First, bearing in mind that
$k_{389}=k_{1083}\times 1083/389$, it is obvious that for equal
saturation parameter and detuning the damping coefficient increases by
a factor 7.8 as compared to a 1083 nm MOT. Although this does not
alter the temperature in the MOT, which does not depend on wavelength
and is expected to be almost equal for the two cases, the damping time
$\tau$ is shortened to $0.13 \tau_{1083}$ \cite{Lett}. Second, and
under the same assumptions for the MOT parameters, the spring constant
is increased by a factor 2.8. This has implications for the size of
the trapped cloud, which is determined by the equipartition of
potential and kinetic energy. The volume $V$ of the cloud is
(following the definition of $V$ as given in Sec.~\ref{cloudshape})
\begin{equation}
\label{MOTvolume}
V=\left(\frac{2\pi k_BT}{\kappa}\right) ^{3/2},
\end{equation}
where for simplicity we have assumed an isotropic 3D harmonic
oscillator ($k_B$ is Boltzmann's constant). It follows that the
volume decreases by a factor 4.5, i.e. the cloud is compressed with
respect to the 1083 nm situation.
\subsection{\label{loading}Loading the MOT}
\begin{figure}[t]
\includegraphics[scale=0.80]{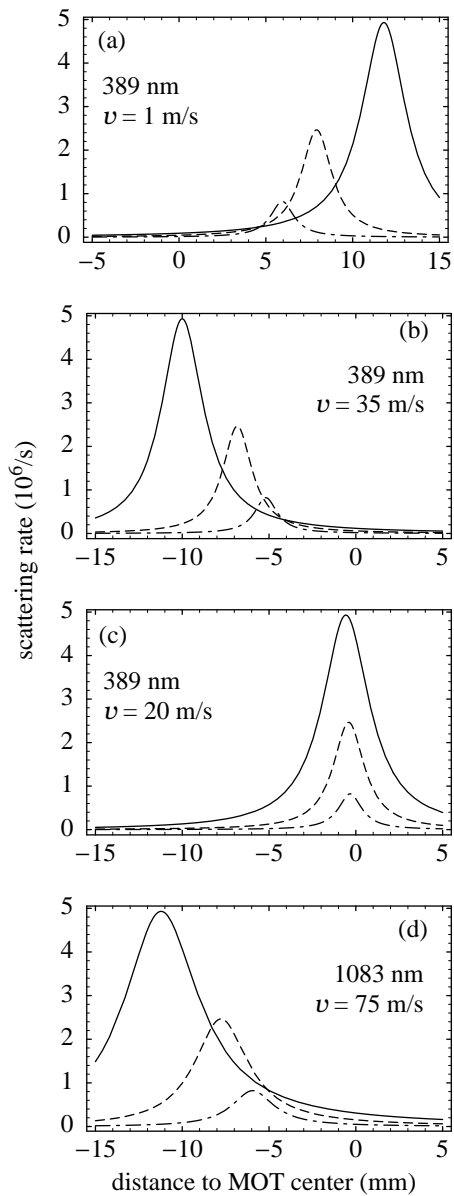}
\caption{\label{MOTrate}(a)-(c) 389 nm and (d) 1083 nm photon scattering 
rates as a function of distance from the MOT center for
$M=1\rightarrow M'=2$ (solid line), $M=0\rightarrow M'=1$ (dashed
line), and $M=-1\rightarrow M'=0$ (dash-dotted line) transitions.}
\end{figure}
All magnetic substates participate in the atom-laser interaction,
since the magneto-optically trapped cloud is contained at low magnetic
field strengths and irradiated from six directions with circularly
polarized light. Therefore, the presence of the second decay channel
of the $3\text{ }^3P_2$ state will not limit operation of the MOT, as
long as there is loading of atoms from the outer regions of the MOT
volume. Loading, however, may be frustrated by the nonclosed cycling
transition as well as by the relatively large Doppler shift. More
specifically, the question arises whether the slowing process of atoms
entering the MOT volume can be completed before a spontaneous emission
via the $3\text{ }^3S_1$ cascade takes the atom to a different,
nonresonant magnetic substate. If not, the atom needs to be repumped
to the cycling transition; otherwise it will escape from the MOT
volume. In order to answer this question, and to obtain a general
notion of the loading process and its dependence on the MOT
parameters, we employ a simple 1D model for an atom traversing the MOT
volume. We calculate the instantaneous photon scattering rate for
atoms at a given velocity $v$, interacting with a counterpropagating,
red-detuned laser beam at 389 nm inducing $\sigma^+$ transitions. This
laser beam represents the two MOT laser beams counterpropagating the
atomic beam at angles of $\pm 45^\circ$ with respect to the atomic
beam (see Sec.~\ref{apparatus}). We assume the atoms to be
predecelerated by a Zeeman slower, so that we can choose any initial
velocity. We take Zeeman detuning, laser intensity, and Doppler shift
into account, the latter of which is taken to be $k v/ \sqrt{2}$ to
correct for the $\pm 45^\circ$ angle between the atom and (real) laser
beams. Furthermore, we consider all three $\sigma ^+$ transitions,
i.e. $M=-1,0,+1
\rightarrow M'=0,+1,+2$ (of which the $M=+1 \rightarrow M'=+2$ will be
referred to as the laser cooling or cycling transition). In
Fig.~\ref{MOTrate}, plots are shown of the photon scattering rate for the 
three $\sigma^+$ transitions as a function of the distance from the center
of the MOT, measured along the symmetry axis of the Zeeman
decelerator. The MOT light boundaries are at about $\pm 10$
mm from the MOT center (see also Sec.~\ref{lasersetup}), and the atoms
are moving into the positive direction. Figure \ref{MOTrate}a shows the
familiar behavior of the scattering force in a MOT. An atom, moving
into the positive direction at a typical intra-MOT velocity $v=1$ m/s,
scatters an increasing number of photons from the counterpropagating
MOT laser beam as it moves farther away from the MOT center.
Consequently, it will be slowed down and eventually pushed back
towards the center.

First, we use this model to investigate the capture of atoms,
emerging from the Zeeman slower in the $2\text{ }^3S_1,M=+1$ state
with a velocity $v=75$ m/s. We choose a MOT detuning of $-35$ MHz, an
intensity of $30I_0$, and a magnetic field gradient of 20 G/cm. These
conditions are typical for a 1083 nm MOT. The model shows that the
resonance condition is never fulfilled inside the MOT volume, thus
preventing any loading of atoms. Next, we lower $v$ to 35 m/s. We
observe that the atoms now interact strongly with the laser light
within the MOT volume (see Fig.~\ref{MOTrate}b). However, the peaks in
the scattering rate of the different $\sigma^+$ transitions hardly
overlap in space, as a result of their different Zeeman detunings.
Slowing these atoms down to zero velocity requires about 190
absorption-emission cycles, whereas it takes about 20 cycles
(corresponding to a velocity reduction of only 4 m/s) for the atom to
end up in one of the nonabsorbing ($M=-1$ or $M=0$)
states. Consequently, the capture process is interrupted. Before this
$M$-state atom becomes sufficiently resonant again, such that it is
optically pumped back to the $M=+1$ state, it will have travelled out
of resonance with the cycling transition and can no longer be captured
by the MOT. Only for velocities $v\le 20$ m/s, an atom ending up in
the wrong $M$ state is repumped fast enough to continue the
deceleration towards zero velocity (Fig.~\ref{MOTrate}c). Thus we
conclude that the capture velocity of the 389 nm MOT is approximately
24 m/s. This velocity is much smaller than the $\sim 75$ m/s capture
velocity of a typical 1083 nm MOT. Figure \ref{MOTrate}d illustrates
the superior loading capabilities of a 1083 nm MOT of 15 mm
radius. The smaller Doppler shift allows for faster atoms to be
captured, whereas the closed cycling transition does not impose any
constraints on the magnetic field strength. In fact, the 1083 nm MOT
diameter sets the maximum stopping distance, and thus limits the
capture velocity. Within the picture provided by the model,
increasing the diameter of a 389 nm MOT will not solve the problem
described above. To avoid optical pumping to nonresonant magnetic substates 
in the outer
regions of the MOT, only small magnetic field gradients can be
tolerated. Then, to maintain sufficient confinement of the trapped
atoms, only small MOT laser detunings are allowed, thereby limiting
the capture velocity. We stress that the model is based on crude
simplifications and ignores important features of the MOT. For
instance, the orthogonal MOT laser beams, in combination with the
spatially varying, three-dimensional magnetic field vector induce
$\sigma$ as well as $\pi$ transitions. Therefore, the conditions
required for repumping to the laser-cooled state may be less stringent
than predicted by our simple model. So we conclude
that a 389 nm MOT will be able to capture sufficiently slow 
metastable helium atoms.

The smaller capture velocity of a 389 nm MOT is a
significant limitation, since a helium atomic beam expands
dramatically due to transverse heating during Zeeman
deceleration~\cite{Minogin}. Calculations of the rms size of the atomic beam
along the slowing trajectory show an increase of the rms atomic beam diameter
by a factor 1.7 when tuning the end velocity from 75 m/s down to
25 m/s. This may lead to a decrease of a factor 3 in metastable flux.
In conjunction with the limited MOT volume, this inevitably will result in a
reduced loading rate.
\section{\label{experimental}Experimental Setup}
\subsection{\label{apparatus}Atomic beam apparatus}
\begin{figure}
\includegraphics[scale=0.55]{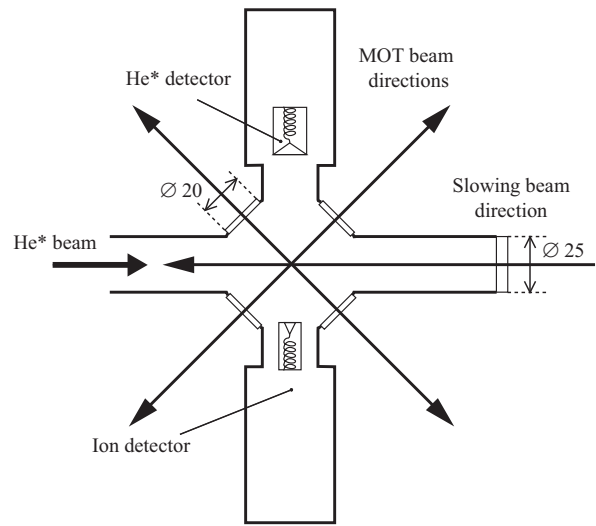}
\caption{\label{MOTchamber}Top view of the MOT vacuum chamber. Not shown are the vertical
MOT laser beams. Dimensions are given in mm.}
\end{figure}

The first stage in our atomic beam apparatus involves a liquid nitrogen
cooled He* DC discharge source, producing an atomic beam that is laser
collimated using the curved-wavefront technique. The beam source is a copy of
the source described by Rooijakkers \textit{et al.} \cite{Rooijakkers}. The
collimated beam enters a differentially pumped two-part Zeeman slower that
reduces the longitudinal velocity from $1000\text{ m/s}$ to $\sim 25\text{
m/s}$. 1083 nm laser light from a commercial 2W fiber laser (measured
bandwidth 8 MHz) is used for slowing and collimation. The laser is stabilized
to the $2\text{ }^3S_1 \rightarrow 2\text{ }^3P_2$ transition using saturated
absorption spectroscopy in an rf-discharge cell. The $-250\text{ MHz}$
detuning for Zeeman slowing
is obtained using an acousto-optical modulator. Downstream the Zeeman slower the MOT vacuum
chamber is located, with 20 mm diameter laser windows for the MOT beams (see
Fig.\ref{MOTchamber}). Two
channeltron electron multipliers are mounted inside to separately
detect ions and metastables. Both channeltrons are
operated with negative high voltage at the front end; however, one of them is put more
closely to the cloud, thereby attracting all positively charged particles and 
leaving only the neutral metastables to be detected by the other. Also, the 
detector of metastables is
hidden behind an aperture in the wall of the vacuum chamber (Fig.~\ref{MOTchamber}), which provides
additional shielding of its electric field. Two 50 A coils,
wound around the vacuum chamber and consisting of 17 turns copper
tubing each, produce a quadrupole magnetic field with a gradient of 43 
G/cm along the symmetry axis. The field of the 
second part of the Zeeman slower inside the MOT region is counteracted with a 
compensation coil, mounted at the position of the Zeeman slower exit. The pressure in the MOT
chamber is $2\times 10^{-9}\text{ mbar}$, and increases to $1\times 10^{-8}\text{ mbar}$ when
the He* beam is switched on.

To minimize atomic beam expansion at the end of the Zeeman slower, we overlap the Zeeman
slower laser beam with an additional 1083 nm beam, with identical circular polarization
and similar intensity, but different detuning ($\Delta =-80\text{ MHz}$). By choosing
the same sign of the quadrupole magnetic field gradient along the Zeeman slower axis as
that of the Zeeman slower itself, an auxiliary Zeeman slowing stage only
centimeters upstream of the MOT volume is established. This should allow trapping of 
atoms with velocities up to $75\text{ m/s}$ at the end of the Zeeman slower. A calculation of
the atomic beam diameter for this case indicates that the loading rate may be
increased by a factor $2.4$ compared to the case where the Zeeman slower decelerates 
atoms to a velocity of $24\text{  m/s}$.
\subsection{\label{lasersetup}389 nm Laser setup}
The MOT laser light is obtained by frequency
doubling the output of a Coherent 899 Titanium:sapphire (Ti:Sa) laser (778 nm with
few-hunderd kHz bandwidth) in an enhancement cavity containing a 10 mm Brewster-cut
LBO crystal. The cavity length is locked to the fundamental wavelength using
the H\"ansch-Couillaud scheme. The Ti:Sa laser is pumped by 10 W at 532 nm
from a Spectra-Physics Millennia X laser. We routinely produce 700 mW of 389 nm
light; peak values of over 1 W of 389 nm at 2.1 W fundamental
power have been achieved. We measured 4\% short-term ($\sim 10$ ms) power fluctuations
in the 389 nm output~\cite{sd}. The LBO crystal is flushed with oxygen, which increases 
the output power by about $10\%$. A small portion
of the UV output is used to stabilize the wavelength to the $2\text{ }^3S_1
\rightarrow 3\text{ }^3P_2$ transition with saturated absorption spectroscopy,
while Zeeman-tuning the Lamb dip allows continuous adjustment of the detuning
between $0$ and $\pm 230\text{ MHz}$. A combination of cylindrical and
spherical lenses transforms the UV beam into a round, parallel and
approximately Gaussian beam with 8 mm waist. The beam profile is truncated by
a 20 mm circular aperture, followed by a series of nonpolarizing beamsplitters
that split the UV beam into four beams. The individual
beam intensities are chosen such that two beams in the horizontal plane can be
retrorefelected, while the intensity of the two vertical beams along the
symmetry axis of the quadrupole field ensures a more or less spherical
He* cloud.
\subsection{MOT diagnostics}
\subsubsection{\label{TOF}Time-of-flight measurement}
The internal energy of helium metastables can be exploited in measuring 
time-of-flight
(TOF) spectra of a MOT. Electron multipliers directly detect
part of the expanding cloud after the atoms in the MOT have been released by
suddenly switching off the MOT laser, the magnetic coils, and the slower
beams. In our experiment, operating the channeltrons in current mode yields
TOF signals resembling a Maxwell-Boltzmann atomic velocity distribution, from
which we can deduce the temperature and number of trapped atoms in the cloud.
However, this procedure requires knowledge of the channeltron gain which
varies with the rate of detected particles and obscures the shape of the TOF
spectrum to an unknown extent. To circumvent an elaborate calibration
procedure, we count the individual metastables that hit the channeltron. This
is, using a properly set amplifier/discriminator, not dependent on the
momentary gain. The output of the amplifier/discriminator is subsequently 
integrated by a calibrated ratemeter. In this way we obtain the total number
of detected metastables during the TOF. Knowing the solid angle covered by the 
detection area, the accuracy in the number of trapped atoms
is now determined by the detection efficiency of
a low-velocity triplet helium atom, which is estimated to be in the range $10-70\%$
(see also Refs.~\cite{Tol,Rooijakkers},
and references therein). This measuring method therefore cannot provide better
than 50\% accuracy in the number of trapped atoms.
\subsubsection{\label{cloudshape}Fluorescence detection}
In addition to the determination of the MOT atom number by time-of-flight 
measurements, we
monitor the fluorescence of the cloud using a calibrated CCD camera to 
independently determine the number of atoms. Here, the cascade via the
$3\text{ }^3S_1$ state offers the 706 nm wavelength, which is far more 
efficiently detected by a camera than fluorescence from a 1083 nm MOT. 
Moreover, the 706 nm light does not suffer from reabsorption, because of the
insignificant population of the $2 \text{ }^3P_2$ level. Therefore we can 
safely assume the monitored fluorescence to be proportional to the number of
atoms at each point in the cloud image, even at the highest densities obtained
in our MOT. To calibrate the camera, we use a small fraction of the
Ti:Sa laser output, with the laser tuned to 706 nm. In the atom
number determination we use dichroic mirrors to block all other wavelengths 
scattered from the MOT, most importantly the abundant 389 nm light. To extract 
the number of atoms $N$ from the observed fluorescence power $P_{fluor}$ we use
the empirical equation of Townsend {\it et al.}~\cite{Townsend}, that
relates the emitted power to the number of atoms: 
\begin{equation}
\label{fluorpower}
P_{fluor}=N\hbar\omega\frac{\Gamma}{2}\frac{6CS}{1+6CS+4\delta^2}.
\end{equation}
In the above equation, $S= I/I_0$, where $I_0$ is the saturation intensity in
the case of $\sigma^+$ transitions in an optically pumped environment, and $I$ 
is the laser intensity of a single MOT beam. The phenomenological factor $C$
incorporates the effects of reduced saturation; as the six circularly polarized
MOT laser beams traverse the cloud in different directions and at varying
angles with the quadrupole magnetic field, all transitions between the ground-
and excited-state Zeeman levels must be considered, and the saturation
intensity $I_0$, as defined above, no longer applies.
It is pointed out in Ref.~\cite{Townsend} that $C$ lies
somewhere halfway the average of the squared Clebsch-Gordan coefficients of 
all involved transitions, and 1. For
the $2\text{ }^3S_1 \rightarrow 3\text{ }^3P_2$ 389 nm transition, the average 
of the squares of the Clebsch-Gordan coefficients is 0.56.
Therefore, we adopt $C=0.8\pm 0.2$, as also chosen by Browaeys 
\textit{et al.}~\cite{OrsayMOT}.
This value incorporates a realistic estimate and
an uncertainty that covers the range of all physically possible values of $C$.

The fluorescence image of the cloud is also used to determine the volume of the
cloud. From a fit to a Gaussian distribution, we obtain the rms size in the
radial ($\sigma_{\rho}$) and axial ($\sigma_z$) directions, and the volume
$V=(2\pi)^{3/2}\sigma_\rho^2 \sigma_z$ ($V$ contains 68\% of the atoms).
For a cloud with Gaussian density distribution, this definition of $V$
conveniently connects the number of atoms $N$ to the central density $n_0$ via
$N=n_0V$. This provides all necessary information to deduce the density
distribution $n({\bf r})$.

\subsubsection{\label{iondetection}Ion detection}
In the MOT vacuum chamber, positive ions are produced in Penning-ionizing 
collisions of a He* atom with
another He* atom or with a background gas molecule.
These ions are subsequently attracted to and detected by the second channeltron,
and the resulting output current provides a rough measure of the number of 
trapped atoms. This signal is particularly useful for optimization purposes.
Moreover, the signal is used to monitor the trap decay after the
loading of the MOT has suddenly been stopped (see Sec.~\ref{traplosses}). 
This channeltron is operated at a sufficiently low voltage, such that the output 
current can safely be assumed to vary linearly with the detection rate.
\section{\label{results}results and discussion}
\subsection{MOT results}
\subsubsection{Temporal fluctuations in the MOT}
While observing the fluorescing cloud in real time with the CCD camera we
noticed nonperiodic intensity fluctuations on a 50 ms timescale. Also, the cloud
was irregularly 'breathing'. To determine
the source of these fluctuations, we first took a series of ten pictures of 
the cloud. The shutter time for each picture was $1/60$ s, and the elapsed
time between two subsequent exposures about 5 s. Fitting the cloud size for 
each individual picture we obtain an average MOT volume with a standard 
deviation of 9\%, while the temperature remained constant within 2.5\%. According to
Eqs.~\ref{omega} and \ref{MOTvolume}, this may be related to the unstable laser 
power. In that case the resulting density fluctuations should influence 
the rate at which ions are produced in two-body Penning collsions. To observe
this, we compared the continuous ion signal with the laser intensity as a
function of time. It turns out that the 4\% laser intensity noise correlates
to the ion signal noise, though it does not explain all irregularities in the 
ion signal. Using Eq.~(\ref{MOTvolume}) we find that the
measured intensity fluctuations may give rise to 6\% variations in the deduced MOT
volume.
\subsubsection{Atom number and density distribution}
The maximum number of loaded atoms as derived from the fluorescence is 
$2.5(3) \times 10^7$ at a detuning $\Delta=-35\text{ MHz}$ and gradient $\partial
B/\partial z=39 \text{ G/cm}$. The total intensity in this case is about $100I_0$. It is
possible to run the MOT at intensities as low as $40I_0$, although the number
of trapped atoms increases with intensity. To ensure a reliable estimate of the cloud
dimensions and fluorescence intensity we take the average of five subsequent images.
The uncertainty in the number of trapped atoms mainly arises from the inaccuracy of the value of the
phenomenological constant $C$ ($8\%$), as well as from an error in the
fluorescence measurement. The uncertainty in the fluorescence measurement is set by the
4\% inaccuracy in the calibration and by the shot-to-shot fluctuations between
the individual images used in the average. To ensure consistency between the
results of the fluorescence and TOF measurement, we have to assign a value of
15(2)\% to the detection efficiency of the channeltron.
A Gaussian density function fits well to the cloud image. From the
fit we infer the rms radii in the $z$ and $\rho$ dimensions and, thus, the
volume $V$. At optimized trapped atom number, we find $V=0.020(5)\text{ cm}^3$. By increasing
the magnetic field gradient to $\partial
B/\partial z=45 \text{ G/cm}$, and decreasing the detuning to $\Delta=-35\text{ MHz}$, the
cloud was compressed to $V=0.0043(4)\text{ cm}^3$. Still, it contained $1.7(2) \times 10^7$ atoms.

Compared to a 1083 nm MOT, typical values for the volume $V$ of the 389 nm MOT are found to
be 6 to 25 times smaller~\cite{Tol}. Although
the auxiliary laser beam at 1083 nm acts as a seventh MOT beam, its effect on the cloud
volume is negligible on account of its large detuning (80 MHz), and the relatively small
photon momentum of the 1083 nm light. Using 
Eq.~(\ref{omega}) and Eq.~(\ref{MOTvolume}), $V$ can be corrected for the
different magnetic field gradients, saturation parameters, and temperatures
for the 389 nm and 1083 nm cases.
It follows that the observed compression of the cloud, due to only the increased
laser cooling force, is approximately a factor 5, as predicted
in Sec.~\ref{springdamping}. The optimum number of atoms is achieved with a relatively
large magnetic field gradient, about twice as large as in a 1083 nm MOT.

With the knowledge of $N$ and $V$ we can determine the central density $n_0
= N/V$, which is $1.4(5)\times 10^9\text{ cm}^{-3}$ in the case of optimized trapped atom number. 
The large error bar, indicating the spread about the mean of the central
densities obtained from each picture, is probably due to the correlation between the volume and the
389 nm laser power fluctuations. A sudden increase in power leads to a smaller volume, while
the fluorescence intensity increases, resulting in an overestimate of the trapped atom number.
The aspect ratio $\sigma_z/\sigma_{\rho}$ of the cloud
turns out to be 0.96(2). We compared this with the aspect ratio as predicted by
Eq.~(\ref{omega}): since at equilibrium $k_BT=\kappa_{\rho}\langle
\rho^2\rangle=\kappa_z\langle z^2\rangle$, with $\kappa_{\rho}$ and $\kappa_z$ the spring
constants of the MOT in the radial and axial directions, respectively, it follows that
$\sqrt{\kappa_{\rho}/\kappa_z}=\sigma_z/\sigma_{\rho}$, resulting in an aspect
ratio of 0.79. This may indicate a small temperature difference between the $\rho$-
and $z$-directions, also observed in a 1083 nm MOT~\cite{Tol}.

\subsubsection{Temperature}
\begin{figure}
\includegraphics[scale=0.80]{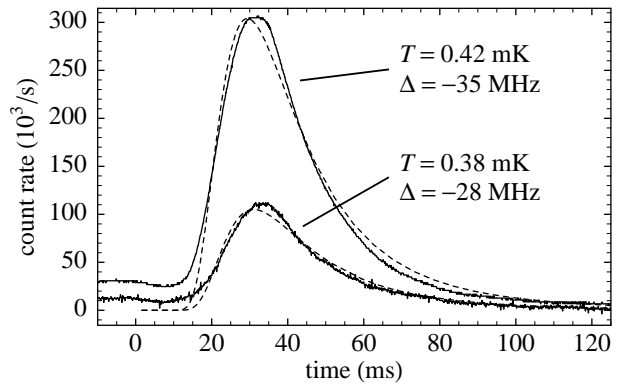}
\caption{\label{TOFsample}Two typical TOF spectra (solid curves) and corresponding fits to the data
(dashed curves), at detunings $\Delta=-35$ MHz and $\Delta=-28$ MHz, respectively. The nonzero offset at
$t\leq 0$ ms is ascribed to loss of metastables during loading of the
MOT, due to imperfect alignment.}
\end{figure}
Fitting a Maxwell-Boltzmann distribution function to the TOF spectra
reveals the temperature $T$ of the atoms in the MOT (Fig.~\ref{TOFsample}).
The fit is not perfect and the deduced temperature may be somewhat overestimated.
Furthermore, a nonzero offset at $t\le 0$ is observed which becomes more prominent
(at the expense of trapped metastables) when
the MOT laser beams are misaligned. The offset may also incorporate 
loss of metastables due to radiative escape~\cite{Weiner}, but our setup does not allow us to
discriminate between different sources of hot metastables.
Measured temperatures range from 0.46(1) mK for $\Delta=-41$ MHz and $S=19$, to
0.25(1) mK at $\Delta=-9$ MHz and $S=15$. In the latter case, however, the number of 
atoms in the MOT is limited to only $2.2\times 10^5$. To our
knowledge, these temperatures are significantly lower than any previously reported 1083 nm
MOT temperature, and even surpass temperatures obtained in 1083 nm 3D-optical
molasses applied to a large cloud of He* atoms~\cite{Tol,OrsayMOT,ENSpobec,
Kumakura,Mastwijk,ANUMOT}.

Commonly, temperatures observed in a 1083 nm MOT are in good agreement with
predictions given by the usual Doppler model~\cite{Lett}:
\begin{equation}
\label{Dopplertemp}
k_BT=-\frac{\hbar\Gamma}{4}\frac{1+2\mathcal{N}S+(2\delta)^2}{2\delta}.
\end{equation}
Here, $\mathcal{N}$ is the dimensionality of the molasses. When using Eq.~(\ref{Dopplertemp}) to
calculate the 389 nm molasses temperature in order to test our results,
two features that distinguish the 389 nm transition from
the 1083 nm transition are relevant. First, the transition strength, determined by the Einstein
coefficient $A_{389}\equiv\Gamma_{389}=2\pi \times 1.5$ MHz, is slightly less than for the 1083 nm
transition ($\Gamma_{1083}=2\pi\times 1.6$ MHz)~\cite{Drake}. This decreases the 389 nm molasses temperature
by 8\% (here $\Gamma_{389}$ should not be confused with the inverse lifetime: 
$(\Gamma_{389})^{-1}=106$ ns,
whereas the lifetime of the $3\text{ }^3P$ state is 95 ns due to the presence of the
extra $3\text{ }^3P \rightarrow 3\text{ }^3S$ decay channel~\cite{Drake}). Second, the 10\% decay via the
$3\text{ }^3S_1$ cascade slightly reduces the diffusion, as the recoil of the photons involved is
randomly distributed. A recalculation of the momentum diffusion constant for this case yields
a 3\% reduction. Thus, we expect the 389 nm molasses temperature to be 11\% lower with respect to the
1083 nm case. Still, the predicted temperature is $1.1\text{ mK}$ for $\Delta=-41$ MHz and $S=19$,
and $0.38\text{ mK}$ for $\Delta=-9$ MHz and $S=15$, appreciably higher than measured under the
same circumstances.

The low temperatures found in our MOT may indicate that
(sub-)Doppler cooling mechanisms are more efficient using the 389 nm
transition. The sub-Doppler frictional force is indeed proportional to
$k^2$. However, the sub-Doppler temperature limit (at low saturation parameter) is
independent of $k$, as the diffusion coefficient also scales with
$k^2$. Moreover,
the capture velocity for such mechanisms is proportional to the wavelength of the transition and
therefore smaller in the 389 nm case.
Also, the recoil temperature associated with the absorption and emission of single 389 nm
photons ($32\text{ }\mu$K) is close to the Doppler limit ($36\text{ }\mu$K), and therefore sub-Doppler
cooling theory is expected to fail for the temperatures observed in our MOT.
At present, the efficiency of sub-Doppler mechanisms in 389
nm molasses compared to 1083 nm molasses remains an open question.

\subsection{\label{traplosses}Trap losses}
\begin{figure}[b]
\includegraphics[scale=0.80]{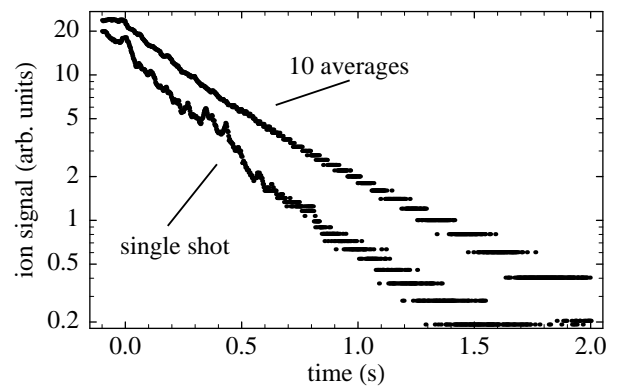}
\caption{\label{decaypic}Lower curve: typical nonexponential decay of the ion signal after
the loading has been stopped at $t=0\text{ ms}$. Upper curve: ion signal obtained after averaging
over 10 decay curves.}
\end{figure}
The number of atoms $N$ in the MOT is governed by the well-known rate equation
\begin{equation}
\label{loadingrate}
\frac{dN(t)}{dt}=L-\alpha N(t)-\beta\int n^2({\bf r} ,t)\,d^3r
\end{equation}
where $L$ denotes the loading rate, and $\alpha$ and $\beta$ are loss rate
coefficients for processes involving one and two metastables,
respectively. Accordingly, when the loading is interrupted the local density
$n$ changes in time following
\begin{equation}
\label{densitylossrate}
\frac{dn}{dt}=-\alpha n-\beta n^2.
\end{equation}
Assuming a Gaussian density profile characterized by a time-independent width,
the losses can be expressed in terms of the central density $n_0$~\cite{Bardou}:
\begin{equation}
\label{centraldensitylossrate}
\frac{dn_0(t)}{dt}=-\alpha n_0(t)-\frac{\beta}{2\sqrt{2}} n_0^2(t).
\end{equation}
The losses are largely due to Penning-ionizing collisions, which yield one
positively charged ion per loss event. These ions are attracted towards the ion
detector, resulting in an ion flux $\phi$.

The loss rate constants are determined from the trap decay when
the loading is stopped by simultaneously
blocking all 1083 nm laser beams entering the apparatus. This disables the
Zeeman slower and collimation section, and prevents the auxiliary Zeeman
slower laser beam of contributing to the two-body collision rate via
light-assisted collisions.
Switching off the collimation minimizes the Penning
ionization contribution of metastables from the atomic beam and, thus, reduces the background
signal.
\subsubsection{\label{decayfit}Collisional losses}
The decay of the MOT is observed by recording the current $\phi(t)$ from the
ion-detecting channeltron~\cite{Tol,Bardou}:
\begin{equation}
\label{ioncurrent}
\phi(t)=V\left( \epsilon_a \alpha n_0(t)+\frac{\epsilon_b\beta}{4\sqrt{2}}n^2_0(t)
\right) +B.
\end{equation}
Here, B is a constant background signal and $\epsilon_a$ and $\epsilon_b$ are the
efficiencies with which ions are produced and detected for losses due to background
and two-body collisions, respectively. Collisions that do not lead to Penning
ionization but do result in trap loss, e.g. collisions with groundstate helium atoms,
reduce $\epsilon_a$. Radiative escape may affect $\epsilon_b$. For the fit procedure,
the ratio $\epsilon=\epsilon_b/\epsilon_a$ must be known. From the increase in
background pressure when the helium atomic beam is running, we deduce that 
the background gas consists for 80\% of helium when the MOT is on.
Unfortunately, our setup is not suited for experimental determination of $\epsilon$, as 
done by Bardou \textit{et al.}~\cite{Bardou}. They experimentally found
$\epsilon=4\pm 1$. Since in our case the background gas involves mainly 
ground-state helium atoms, we expect $\epsilon_a$ to be smaller than unity. The value of $\epsilon_b$ is
probably close to unity: following Tol \textit{et al.}~\cite{Tol}, one finds that for the 1083 nm case
$\epsilon_b\approx 0.98$. We take the obvious underestimate $\epsilon=1$,
which implies that the result of the fit for
$\beta n_0$ has to be considered an upper limit. The result for $\alpha$ can also be obtained by fitting the
tail of the decaying ion signal, where the density is low enough to neglect the contribution of the
two-body losses.
In this way, the significance of $\epsilon$ in the determination of $\alpha$ is strongly
(but not completely) reduced.

A typical example of a decaying ion signal is depicted in
Fig.~\ref{decaypic}. The decay clearly shows nonexponential behavior, indicating that two-body collisions
contribute significantly to the total losses.
Since laser power fluctuations cause density fluctuations, much noise is visible in the ion
signal. Therefore an average of ten decay transients is fitted, as also shown in Fig.~
\ref{decaypic}. Unfortunately,
this may affect the reliability of the fitted parameters as the two-body loss rate
depends nonlinearly on intensity. However, apart from intensity noise the 389 nm output remained
constant over a period sufficiently long to perform the measurements.

The fit procedure yields values for the exponential time constant $\alpha$ and the
nonexponential time constant $\beta n_0$. We typically find $\alpha=2\text{ s}^{-1}$
and $\beta n_0=3\text{ s}^{-1}$.
This gives the rate coefficient $\beta$ from the fit parameter
$\beta n_0$, using $n_0$ from the fluorescence measurement. We find
$\beta=1.0(4)\times 10^{-9}\text{ cm}^3/\text{s}$, at a detuning of $-35\text{
MHz}$. Assuming a value $\epsilon = 4$, the result becomes 
$\beta=6(2)\times 10^{-10}\text{ cm}^3/\text{s}$. The value $\beta=1.0(4)\times 10^{-9}\text{ cm}^3/\text{s}$,
which we interpret as the upper limit, is significantly below the
value for the 1083 nm case of
$5.3(9)\times 10^{-9}\text{ cm}^3/\text{s}$, reported by Tol {\it et
al.}~\cite{Tol} using the same detuning and similar saturation.
\begin{table}
\caption{\label{comptable}Comparison of the 389 nm MOT with the 1083 nm MOT described in Ref.
~\cite{Tol}.
The typical results for both MOTs are obtained under conditions that optimize both density and
atom number. For the 389 nm case $\epsilon=1$ is assumed.}
\begin{ruledtabular}
\begin{tabular}{lll}
MOT wavelength& 389 nm& 1083 nm \\
\hline
\\
Detuning $\Delta$ (MHz)& $-35$&$-35$\\
Magnetic field gradient\\
$\partial B/\partial z$ (G/cm) & 41 & 20 \\
Total intensity ($I_0$)& 100&90\\
Number of atoms $N$& $2 \times 10^7$& $5 \times 10^8$\\
Loading rate $L$ (s$^{-1}$)& $<10^8$& $>5 \times 10^9$\\
Central density $n_0$ (cm$^{-3})$& $4\times 10^9$& $4\times 10^9$\\
Volume $V$ (cm$^3$)& 0.005& 0.12\\
Temperature $T$ (mK)& 0.42& 1.1\\
Two-body loss rate $\beta n_0$ (s$^{-1}$)& 3& 21\\
Two-body loss rate constant\\ 
$\beta$ (cm$^3$/s)& $1.0(4)\times 10^{-9}$
& $5.3(9)\times 10^{-9}$\\
Two-photon ionization\\ loss rate
constant $\alpha_{2ph}$ (s$^{-1}$)& 0.5 & 0\\
\end{tabular}
\end{ruledtabular}
\end{table}

The small value for $\beta$ may be explained by a simple argument from
cold-collision theory. A light-assisted
collision can be regarded as two $2\text{ }^3S_1$ atoms that are resonantly excited to a
molecular complex. For small detunings, this occurs at relatively large internuclear separation,
where the molecular potential $U$ is well-approximated by the dipole-dipole
interaction
\begin{equation}
\label{molpot}
U_{\pm}=\pm\frac{C_3}{R^3}.
\end{equation}
Here, $R$ is the internuclear distance, and
$C_3\simeq\hbar\Gamma(\lambda/2\pi)^3$~\cite{Weiner}. The excitation by the red-detuned MOT
laser light takes place resonantly when the molecular potential energy
compensates the detuning. This sets the so-called Condon radius $R_C$:
\begin{equation}
\label{condonpoint}
R_C=\left(\frac{C_3}{2\pi\hbar |\Delta|}\right)^{1/3}.
\end{equation}
The red detuning selects an attractive molecular state. Once excited, the two atoms are
accelerated towards small internuclear distances, where Penning ionization occurs with high probability.
It follows from Eq.~(\ref{molpot}) and Eq.~(\ref{condonpoint}) that the Condon
radius for 389 nm excitation is 2.8 times smaller than for 1083 nm. Classically,
the cross section for the collision is determined by the square of the Condon radius,
and is therefore expected to decrease by almost a factor 8.

To identify the role played by light-assisted collisions in the total two-body losses,
we assume that $\beta$, as defined in Eq.~(\ref{densitylossrate}), can be decomposed in two 
terms: $\beta_{SS}$ and $\beta_{SP}$. Here $\beta_{SS}$ is the rate coefficient for
losses due to collisions between $2\text{ }^3S_1$ atoms in the absence of light, whereas
$\beta_{SP}$ takes the light-assisted collisional losses into account and depends
(for given detuning and saturation parameter) on the cross section and, thus, on the
Condon radius as described above. We neglect collisions between excited-state atoms, since the 
excited-state population in our far-red-detuned MOT does not exceed 0.01. We can define 
$\beta_{SS}$ and $\beta_{SP}$ also via Eq.~(\ref{densitylossrate}), with the total density $n$ 
replaced by the $2\text{ }^3S_1$ density $n_S$:
\begin{equation}\label{groundstatelosses}
\frac{dn_S}{dt}=-\alpha n_S-(\beta_{SS}+\beta_{SP}) n_S^2.
\end{equation}
Since the excited-state population is small, $n_S\approx n$. It now follows immediately
from Eq.~(\ref{densitylossrate}) and Eq.~(\ref{groundstatelosses}) that to good approximation $\beta=\beta_{SS}+\beta_{SP}$.

$\beta_{SS}$ has been measured in a 1083 nm MOT by Tol \textit{et al.}~\cite{Tol} to be
$\beta_{1083}^{SS}=2.6(4)\times 10^{-10}\text{ cm}^3$/s.
Subtracting this from the total rate coefficient $\beta_{1083}=5.3(9)\times10^{-9}\text{ cm}^3$/s,
we infer $\beta_{1083}^{SP}=5(1)\times10^{-9}\text{ cm}^3$/s, which is much larger than $\beta_{1083}^{SS}$.
In contrast, the upper limit we find for $\beta_{389}$ is of the same order
of magnitude as $\beta_{389}^{SS}$
(since 1083 nm and 389 nm magneto-optical traps, operated under the same conditions, are assumed to lead
to similar populations of the 2 $^3S_1$, $M=-1,0,1$ levels, we can take $\beta_{389}^{SS}=\beta_{1083}^{SS}$).
To obtain the upper limit for $\beta_{389}^{SP}$, we subtract $\beta_{389}^{SS}$ from $\beta_{389}$ and find
$\beta_{389}^{SP}\le 7(3)\times10^{-10}\text{ cm}^3$/s. This is in good agreement with the
prediction following our simple argument.
In addition to the upper limit found for $\beta_{389}$, we can now assign a lower limit equal to
$\beta^{SS}_{389} =2.6(4)\times 10^{-10}\text{ cm}^3$/s. Summarizing, we find $2\times 10^{-10}
\text{ cm}^3/\text{s} < \beta_{389}
< 1.0 \times 10^{-9}$ cm$^3$/s.
\subsubsection{Two-photon ionization}
\label{twophoton}
\begin{figure}[b]
\includegraphics[scale=.81]{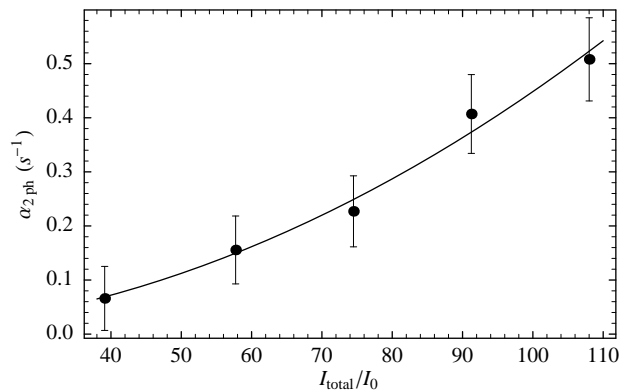}
\caption{\label{alfavssat}Two-photon loss rate constant versus total saturation parameter $I_{total}/I_0$
at a detuning $\Delta=-35$ MHz.}
\end{figure}

From the fit to the ion signal decay, we extract the linear loss rate
coefficient $\alpha$. Unlike the situation in 1083 nm magneto-optical traps,
$\alpha$ is not solely determined by background-gas collisions, but also by
the two-photon ionization rate. We assume that each loss event involves only one
He* atom and ignore photoionization of the molecular complex formed during
a light-assisted collision, as this process enters
Eq.~(\ref{centraldensitylossrate}) via $\beta$. Hence the loss rate coefficient
can be written
\begin{equation}
\alpha=\alpha_{bgr}+\alpha_{2ph},
\end{equation}
\begin{figure}[t]
\includegraphics[scale=.82]{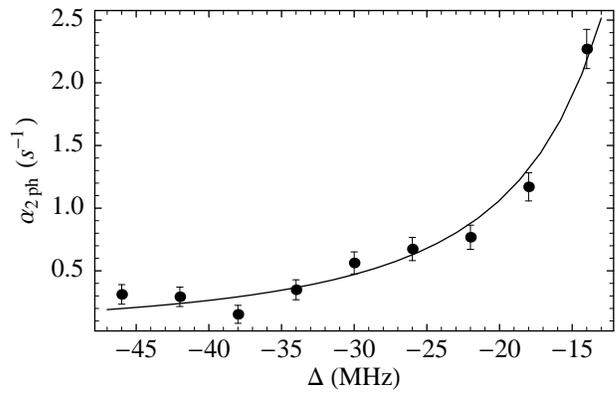}
\caption{\label{alfavsdet}Two-photon loss rate constant $\alpha_{2ph}$ versus
MOT detuning for total saturation parameter $I_{total}/I_0=6S=110$.}
\end{figure}
where $\alpha_{bgr}$ denotes the background gas collisional rate, and
$\alpha_{2ph}$ accounts for the two-photon ionization loss rate. Two processes
can be thought to cause the
ionization: two-photon ionization of a $2\text{ }^3S_1$ atom, or
photoionization of an atom in either the $3\text{ }^3P_2$ or the $3\text{ }^3S_1$ state.
The latter state is populated only during the cascade and has a lifetime of only 35 ns,
so its contribution will be negligible.
The instantaneous two-photon ionization probability $p_{inst}$ is, for not too large
detuning $\Delta$, dependent on intensity $S$ and MOT detuning $\Delta$ according to
\begin{equation}
\label{instion}
p_{inst}\varpropto \frac{S^2}{\Delta^2}.
\end{equation}
The photoionization probability $p_{pi}$ of a helium atom in the $3\text{
}^3P_2$ state is simply proportional to the incident laser intensity and the
cross-section for photoionization, which varies only slowly
with wavelength~\cite{ChangandFang}. Neglecting this wavelength dependence, the
probability of photoionization simply becomes
the product of the upper $3\text{ }^3P_2$ state population and the ionization
probability itself. For the two-step process, this leads to a dependence on
intensity and detuning as
\begin{equation}
p_{pi} \varpropto \frac{S^2(\Gamma/2)^2}{\Delta^2+(S+1)(\Gamma/2)^2}.
\end{equation}
When $\Delta^2 \gg (S+1)(\frac{\Gamma}{2})^2$, this dependence takes on a form similar to
Eq.~(\ref{instion}). We confirmed this behavior by measuring $\alpha_{2ph}$ as
a function of MOT detuning, as shown in Fig.~\ref{alfavsdet}. We also checked the
intensity dependence, as shown in Fig.~\ref{alfavssat}.
In both cases we determined $\alpha_{bgr}$ by measuring $\alpha$ as a
function of background pressure, while keeping the detuning and intensity 
fixed. Assuming a linear variation of $\alpha_{bgr}$ with pressure, against 
a fixed background $\alpha_{2ph}$, a fit to the data points yields 
$\alpha_{bgr}\approx 1.5(1)\text{ s}^{-1}$. Under typical experimental conditions, we find
$\alpha_{2ph}\approx 0.5\text{ s}^{-1}$.

Chang \textit{et al.} calculated photoionization cross sections of many singlet
and triplet states in helium, including the $3\text{ }^3S$ and $3\text{ }^3P$ 
states, for various wavelengths~\cite{ChangandFang}. Using their results we
find photoionization rates of about $2\text{ s}^{-1}$. Since the fraction of 
$n=3$ atoms in our MOT is typically below
the 1\% level, the net loss rate due to the two-step process then would be one order of
magnitude smaller than the measured value for $\alpha_{2ph}$. This suggests
that instantaneous two-photon ionization dominates over the two-step
ionization losses.
\subsection{Auxiliary Zeeman slower, loading rate, and MOT capture velocity}
To test the performance of the auxiliary Zeeman slower, we first optimized the
number of atoms in the MOT in the absence of the extra slowing laser. Then,
leaving the MOT parameters unaltered, we unblock the auxiliary laser beam and
vary the slower laser intensities and Zeeman coil current iteratively until a new
optimum for the number of atoms is established. Indeed, blocking the
additional laser beam again interrupts the loading, demonstrating that we have tuned
the end velocity of the Zeeman decelerator above the capture velocity of the
MOT. With the auxiliary Zeeman slower on,
the number of atoms is 40\% times larger as compared to the case
without the auxiliary Zeeman slower. Making use of Eq.~(\ref{loadingrate}), with
$\alpha$, $\beta$, and $n_0$ known from experiment, we calculate that
the auxiliary Zeeman slower enhances
the loading rate by a factor 1.6.
Despite this improvement, the loading
rate remains low. By solving Eq.~(\ref{loadingrate}), with
the measured values for the loss rate constants and the steady-state number of
atoms as input, we find a loading rate slightly below $10^8 \text{ s}^{-1}$.
Tol \textit{et al.}~\cite{Tol} state a value of $5 \times 10^9\text{ s}^{-1}$ for their 1083
nm MOT. This difference is explained by the smaller MOT diameter, the reduced
flux of slow atoms from the Zeeman slower due to atomic beam expansion, and imperfect collimation due to
the relatively large bandwidth of the 1083 nm laser.

From the Zeeman slower settings it is possible to calculate the end velocity of the atoms and, thus,
the capture velocity of the MOT. Therefore, the equations of motion of an atom subject to the
decelerating laser beam are solved. We take into account 
the saturation parameter, the laser beam intensity profile,
and the magnetic field (obtained from a detailed calculation). In this way we derive a
capture velocity of 35 m/s (without using the auxiliary Zeeman slower).
Apparently, the prediction of a 24 m/s capture velocity by the model of
Sec.~\ref{loading} is an underestimate, and the true capture velocity lies closely to the velocity
determined by the resonance condition. Therefore, it is likely that $\pi$ and $\sigma$ transitions,
caused by MOT laser beams orthogonal to the quantization axis, occur at rates at least comparable to
the 10\% decay via the 3 $^3S_1$ cascade. Apparently, the nonclosed character of the 389 nm
transition plays a minor role, even in the case of relatively large ($\sim 40\text{ }G$) magnetic fields.

We derive from the settings of the Zeeman slower that atoms with a velocity of at
most 75 m/s are further decelerated to a velocity of 35 m/s by the
auxiliary Zeeman slower. This translates to an increase in loading rate by a factor 1.7, in 
reasonable agreement with the result of the test described above.
\subsection{Comparison with 1083 nm MOT}
Table \ref{comptable} contains MOT results for the 389 nm and 1083 nm case~\cite{Tol}. Both MOTs
have similar detuning and saturation parameter, which optimize both density and trapped atom number.
The smaller number of atoms $N$ in the 389 nm MOT is
explained by the small loading rate. Despite this small number, the central density $n_0$
is equal to that of a 1083 nm MOT containing over one order of magnitude more atoms. This is the result of the
smaller loss rate constant $\beta$, the larger laser cooling force, and the larger magnetic field gradient.
The latter not only contributes to the compression of the cloud, but also reflects the necessity
of a large Zeeman detuning to compensate the larger Doppler shift of the atoms to be captured from the
Zeeman-decelerated He* beam. Furthermore, we observe that
the 0.5 s$^{-1}$ contribution of two-photon ionization to the losses in
the 389 nm MOT is small compared to the 21 s$^{-1}$ two-body loss rate in a large 1083 nm MOT.
\section{Conclusion and Outlook}
\label{conclusion}
We have shown that it is possible to build a magneto-optical trap using the 389 nm transition in triplet
helium. Our prototype MOT demonstrates that a 389 nm MOT offers the
advantage of a dense, cold cloud of metastable helium atoms, as compared to
a 1083 nm MOT. The relatively large density is allowed by the reduced
two-body loss rate coefficient $\beta$, whereas the large spontaneous force moderates
substantial compression of the cloud. Intensity noise on the 389 nm output, however, compromises the
measurement accuracy. Together with the high background pressure and the small value of $\beta$, this has
complicated an accurate determination of its value. We conclude that $\beta$ lies
between the experimentally determined upper limit $1.0\times 10^{-9}$ cm$^3$/s, and the two-body loss rate
constant in the absence of light, $2\times 10^{-10}$ cm$^3$/s determined in Ref.~\cite{Tol}.
Two-photon ionization losses, although
present, do not exclude the future possibility of a 389 nm MOT
containing large numbers of metastable helium atoms at high phase-space
density. To this end, however, the loading rate of the MOT must be improved. A bare 389 nm
MOT has limited loading capabilities since
the large Doppler shift implies a reduced capture velocity, and the required Zeeman slower
settings then give rise to a smaller flux of slow metastables. The nonclosed character of the
389 nm transition, however, does not play an important role in the capture process, as well
as in the other physics involved in the MOT.

For the near future we plan an ultimate experiment, in a
configuration with a loading rate increased by two orders of magnitude. To realize this,
a 1083 nm MOT with $\sim 30$ mm diameter laser beams will be overlapped
with a $\sim 10$ mm diameter 389 nm MOT. To avoid large two-body losses in the trapped cloud
due to the presence of red-detuned 1083 nm light, a $\sim 5$ mm diameter hole has to
be created in the center of the 1083 nm MOT laser beams. This configuration will benefit
from the superior loading capability of the 1083 nm MOT, as well as from the low-loss
389 nm environment containing a dense cloud at relatively low temperature. Furthermore,
we will test the effectiveness of 389 nm molasses on a metastable helium cloud, precooled
by a 1083 nm MOT. This seems promising not only because of the low temperatures observed already
in our 389 nm MOT, but also because of the relatively high saturation intensity. This can be
seen as follows. In a 1083 nm molasses, starting with 
a large, dense cloud of helium metastables, the relative absorption is rather large~\cite{ENSpobec}.
Within the cloud, this results in intensity imbalances between two counterpropagating molasses laser beams,
and these imbalances are believed to
reduce the effectiveness of the molasses. The larger saturation intensity of the 389 nm transition implies
lower relative absorption and, thus, a reduced intensity imbalance. Therefore, we expect to obtain lower
temperatures in a 389 nm molasses.

\section{Acknowledgements}
We are indebted to J. Bouma for his contribution to the design and construction of the
setup. We thank P.J.J. Tol and N. Herschbach for stimulating discussions. The Space Research
Organization Netherlands (SRON) is gratefully acknowledged for financial support.
\newpage
\bibliographystyle{apsrev}
\bibliography{articletext}

\end{document}